\documentclass{article}

\PassOptionsToPackage{numbers, compress}{natbib}

\usepackage[final]{neurips_2025_ml4ps}

\usepackage[utf8]{inputenc} 
\usepackage[T1]{fontenc}    
\usepackage{hyperref}       
\usepackage{url}            
\usepackage{booktabs}       
\usepackage{amsfonts}       
\usepackage{nicefrac}       
\usepackage{microtype}      
\usepackage{xcolor}         
\usepackage{amsmath}
\usepackage{amssymb}
\usepackage{graphicx}

\usepackage{natbib}
\title{Unsupervised Discovery of High-Redshift Galaxy Populations with Variational Autoencoders}

\author{%
  Aayush Saxena \\
  Department of Physics\\
  University of Oxford\\
  Denys Wilkinson building, Oxford, OX1 3RH, UK \\
  \texttt{aayush.saxena@physics.ox.ac.uk} \\
}

\begin{document}

\maketitle

\begin{abstract}
  We apply variational autoencoders to automatically discover galaxy populations using publicly available high-redshift \textit{JWST} spectra without prior classification knowledge. Our unsupervised method identifies distinct astrophysical classes of unique and exciting galaxy types, demonstrating automated discovery capabilities for large spectroscopic surveys.
\end{abstract}

\section{Introduction}

Since its launch on Christmas Day in 2021, the \textit{James Webb Space Telescope (JWST)} has rapidly been transforming our understanding of how the first galaxies form and evolve a few hundred million years after the Big Bang, when the Universe was a fraction of its current age of 13.6 billion years. This has mainly been achieved thanks to \textit{JWST}'s imaging and spectroscopic capabilities at near-infrared wavelengths, which trace the redshifted ultraviolet and optical light from distant galaxies. 

Spectroscopy of galaxies is one of the most important tools to analyze their physical and chemical properties and test theories of galaxy formation and evolution. Particularly for distant, high redshift galaxies, the shape of the continuum emission from stars, supermassive black holes and hot gas, and emission lines from nebular regions, can give insights into the dominant sources of photoionization, the physical (temperature, density) and chemical (level of enrichment from elements heavier than Hydrogen and Helium) properties of the stellar populations and the interstellar gas, and their cosmic dust content, which are all vital building blocks of galaxies.

Early \textit{JWST} spectroscopic results have shed new light on several outstanding questions in the field, such as when and how did the first stars and supermassive black holes form in early galaxies, how are the first generation of stars different from evolved stars in our Milky Way, and what is the impact of the radiation emitted by the earliest galaxies on the cosmological evolution of the intergalactic medium. These early results have often relied on the identification of small samples of interesting galaxies `by eye' from individual large observing programs. With over three years of scientific operations and a growing repository of publicly available spectroscopic data from several large and treasury observing campaigns, the need of the hour is to assemble statistically significant samples of the most interesting galaxies to better inform models of galaxy evolution.

Machine learning approaches, particularly unsupervised deep learning models, are perfectly suited to automate the discovery and classification of astrophysical objects at scale. Variational Autoencoders (VAEs; \citep{kingma2013}) can be powerful when applied to spectroscopic data analysis as they learn compact, interpretable representations from intrinsically complex datasets, while enabling both reconstruction and generation of synthetic data. Unlike supervised methods that require a large repository of labeled data, VAEs are capable of discovering structure in the latent space in an unsupervised fashion. VAEs have previously been applied to large spectroscopic datasets of nearby, low redshift galaxies taken from ground-based telescopes \citep[e.g.][]{portillo2020, bohm2023, scourfield2023, nicolau2025}, but have never been deployed in the context of high redshift galaxy spectra from \textit{JWST}, which represents a potent discovery space.

In this work, we leverage VAEs and latent space clustering to discover and characterize statistically significant samples of rare and exciting high redshift ($z>4$) galaxies, probing the first 1.5 billion years of the Universe's evolution. In Section\,\ref{sec:method} we describe the implementation of the VAE, in Section\,\ref{sec:results} we present our main results, in Section\,\ref{sec:future} we discuss future directions for this work, and in Section\,\ref{sec:conclusions}. We have made the datasets and code publicly available\footnote{\url{https://github.com/aayush3009/learnspec}} in the spirit of reproducibility and open science.

\section{Methods}
\label{sec:method}

\subsection{Variational autoencoder (VAE) architecture}
In this work, we implement a Variational Autoencoder (VAE) following the framework developed by \citet{kingma2013}. The VAE learns a probabilistic mapping between high-dimensional astronomical spectroscopic data and a lower-dimensional latent space. The VAE optimizes the Evidence Lower Bound (ELBO): 
\begin{equation}
    \mathcal{L} = \mathbb{E}_{q\phi(z|x)}[\log p_\theta(x|z)] - D_{KL}(q_\phi(z|x) || p(z))
\end{equation}
where the first term represents the reconstruction accuracy and the second enforces regularization towards a prior distribution, typically a Gaussian, $p(z) = \mathcal{N}(0, I)$ via calculation of the Kullback-Leibler (KL) divergence. The encoder neural network, $q_\phi(z|x)$, learns to map the input spectra, $x \in \mathbb{R}^d$ to latent parameters, $(\mu, \sigma^2) \in \mathbb{R}^{2k}$, where $d$ is the input dimension and $k$ is the latent space dimension. The decoder neural network, $p_\theta(x|z)$ reconstructs the original spectra from latent variables, $z \in \mathbb{R}^k$ sampled via the reparameterization trick \citep{kingma2013}.

For the encoder and the decoder, we employ a deep symmetric neural network architecture with four fully-connected layers. For the encoder, the layers progressively compress the dimensionality, $d$ of the input spectra: $d \rightarrow 512 \rightarrow 256 \rightarrow 128 \rightarrow 64 \rightarrow k$, where $k=16$ is the dimensionality of the latent vectors chosen in our implementation. The choice of $k=16$ balances expressiveness of the neural network with computational efficiency. The decoder architecture mirrors that of the encoder, expanding the latent space back to the dimensionality of the input spectra: $k \rightarrow 64 \rightarrow 128 \rightarrow 256 \rightarrow 512 \rightarrow d$.

We implement a range of regularization techniques to prevent overfitting and improve generalization: (i) we apply L2-weight regularization with $\lambda = 0.001$ to all hidden layers; (ii) we apply batch normalization after each dense layer to stabilize training; (iii) we deploy dropout layers in the encoder network with rates decreasing from $0.2$ to $0.1$ towards the latent bottleneck (and vice-versa in the decoder); and (iv) we apply gradient clipping for training stability. For spectral data containing missing/masked inputs, we implement a masked reconstruction loss:
\begin{equation}
    L_{rec} = \frac{1}{N} \sum_{i=1}^N \sum_{j=1}^d M_{ij} (x_{ij} - \hat{x}_{ij})^2
\end{equation}
where $M_{ij}$ is a binary mask that excludes missing spectral data.

To train the networks, we implement exponentially decaying learning rates starting at $10^{-4}$ with a decay rate of $0.95$ every 500 steps, combined with early stopping if the validation reconstruction loss stops improving after 50 steps. The training set contained $85\%$ of the data and the validation set contained $15\%$ of the data. 

\subsection{Data pre-processing}
The spectroscopic data are taken from the DAWN \textit{JWST} Archive\footnote{\url{https://dawn-cph.github.io/dja/index.html}} (DJA), which is a repository containing nearly all publicly available \textit{JWST} datasets. DJA further provides redshift information inferred from the galaxy spectra along with quality flags. In this work we only use redshifts with the highest quality flag, considering only sources with redshifts above 4 ($z>4$; tracing the first 1.5 billion years after the Big Bang). There are 2743 objects in our final dataset.

A number of pre-processing steps must be applied to prepare the datasets before feeding them into the VAE. The first step involves de-redshifting the spectroscopic data to resample the spectra into the rest-frame wavelength ($\lambda_{\rm{rest}} = \lambda_{\rm{obs}}/(1+z))$. A uniformly spaced rest-wavelength grid was determined based on the median redshift of the sample. Since the observed wavelength range of \textit{JWST}/NIRSpec is fixed from $\sim7500-53000$\,\AA, spectra at different redshifts will sample different rest-frame wavelength ranges. Any resulting missing spectroscopic flux on the common rest-frame wavelength grid was masked. 

Each de-redshifted spectrum was then normalized by scaling its continuum flux at rest-frame $1500$\,\AA\ to $1.0$. Spectroscopic data contain strong nebular emission lines and lower flux continuum tracing starlight. Therefore, to properly leverage the discerning power within the dynamic range of fluxes after normalization, we used a novel arcsinh transformation ($\rm{arcsinh}(x) = \ln \left(x + \sqrt{x^2+1}\right)$), which is approximately linear for small values of $x$ (continuum), and log for large values of $x$ (emission lines). This helps preserve information from both the continuum shape and emission lines, which are important features for galaxy the classification task at hand.

\section{Results}
\label{sec:results}

\subsection{Reconstruction accuracy}

\begin{figure*}
    \centering
    \includegraphics[width=\linewidth]{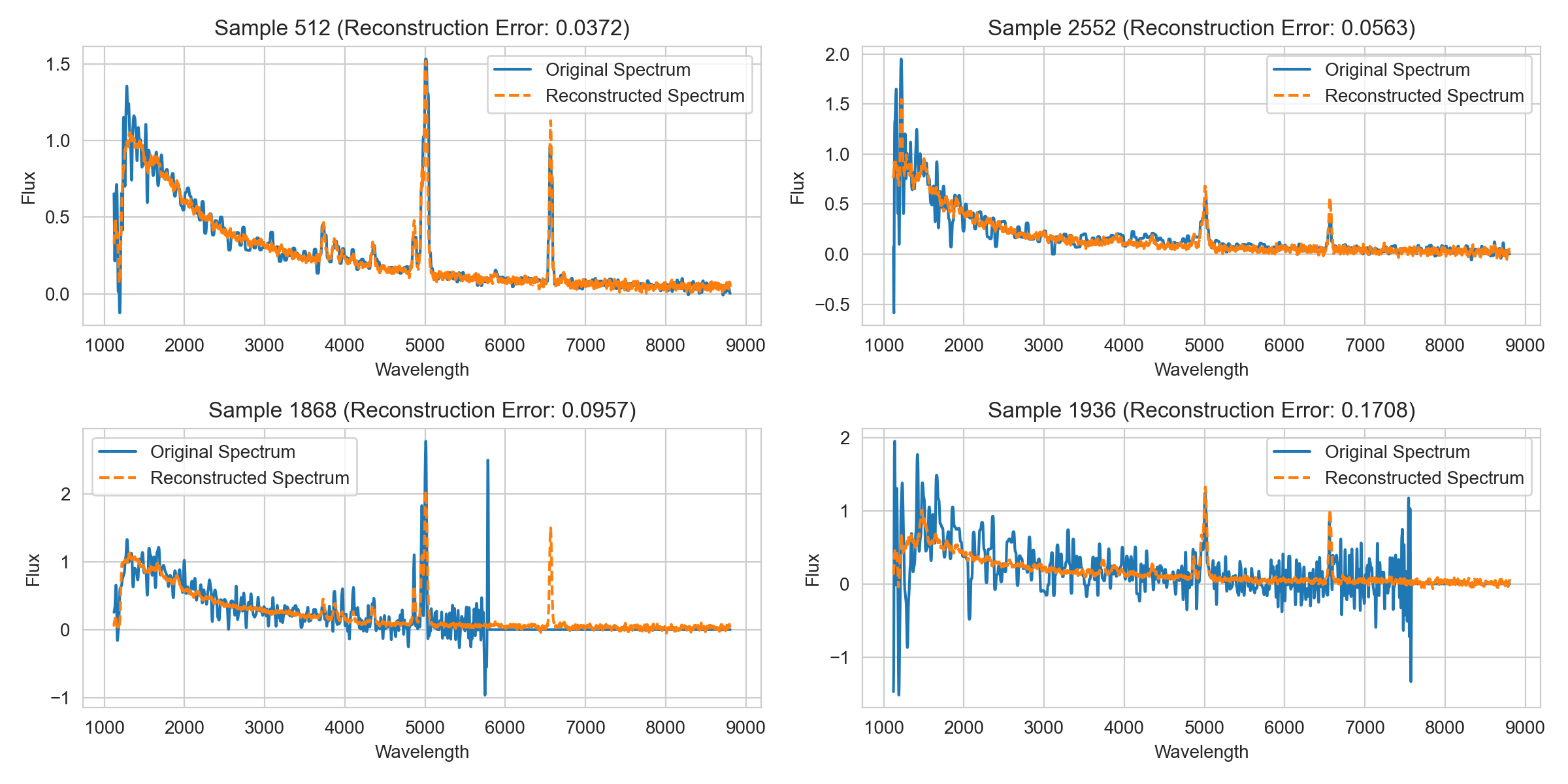}
    \vspace{-6mm}
    \caption{Comparisons of input (blue) and reconstructed (orange) spectra, drawn from four quartiles of reconstruction errors distribution, with decreasing accuracy clockwise from top-left. We note that the reconstruction often makes predictions for when the input data is missing/masked.}
    \vspace{-4mm}
    \label{fig:reconstruction}
\end{figure*}
Our VAE model performs excellently at reconstructing the vast majority of spectra. The reconstruction error distribution measured as the mean squared error (MSE) between the original and reconstructed spectra has a median value of $0.122$ and is one-sided, long-tailed Gaussian with a standard deviation of $0.124$. In Figure \ref{fig:reconstruction} we show representative examples from four quartiles of the error distribution. The VAE is able to reconstruct masked/missing flux that often leads to higher reconstruction errors. High-error reconstructions (MSE $>0.1$) typically trace noisy spectra that contain artifacts, or spectra with extremely faint continua.

\subsection{Clustering in latent space}
To identify interesting galaxy types from the latent space while breaking the `curse of dimensionality', we collapse the 16D latent space to a 2D representation using UMAP dimensionality reduction \citep{mcinnes2019}. We apply Gaussian Mixture Modeling with increasing number of components in the range $[5,15]$ to the 2D embeddings 100 times and record the clustering solution that returns the maximum Silhouette score \citep{rousseeuw1987}. With a Silhouette score of $0.44$, we identify 12 well-separated clusters that include clusters containing noisy and artifact-dominated data. The number of galaxies across clusters ranges from 63 to 334 with no clear dominant class, demonstrating that our model is capable of capturing diverse galaxy populations across redshifts.

\subsection{Astrophysical insights}
\begin{figure*}
    \centering
    \includegraphics[width=0.71\linewidth]{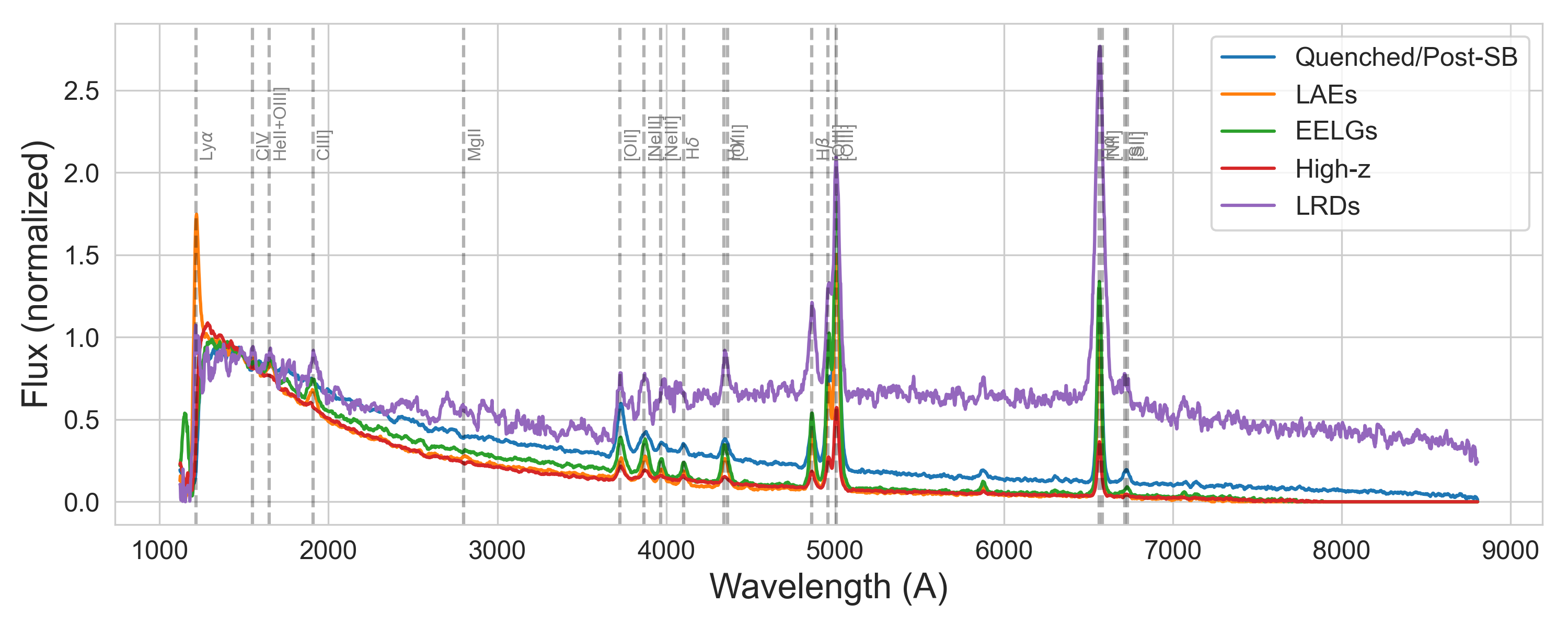}
    \includegraphics[width=0.28\linewidth]{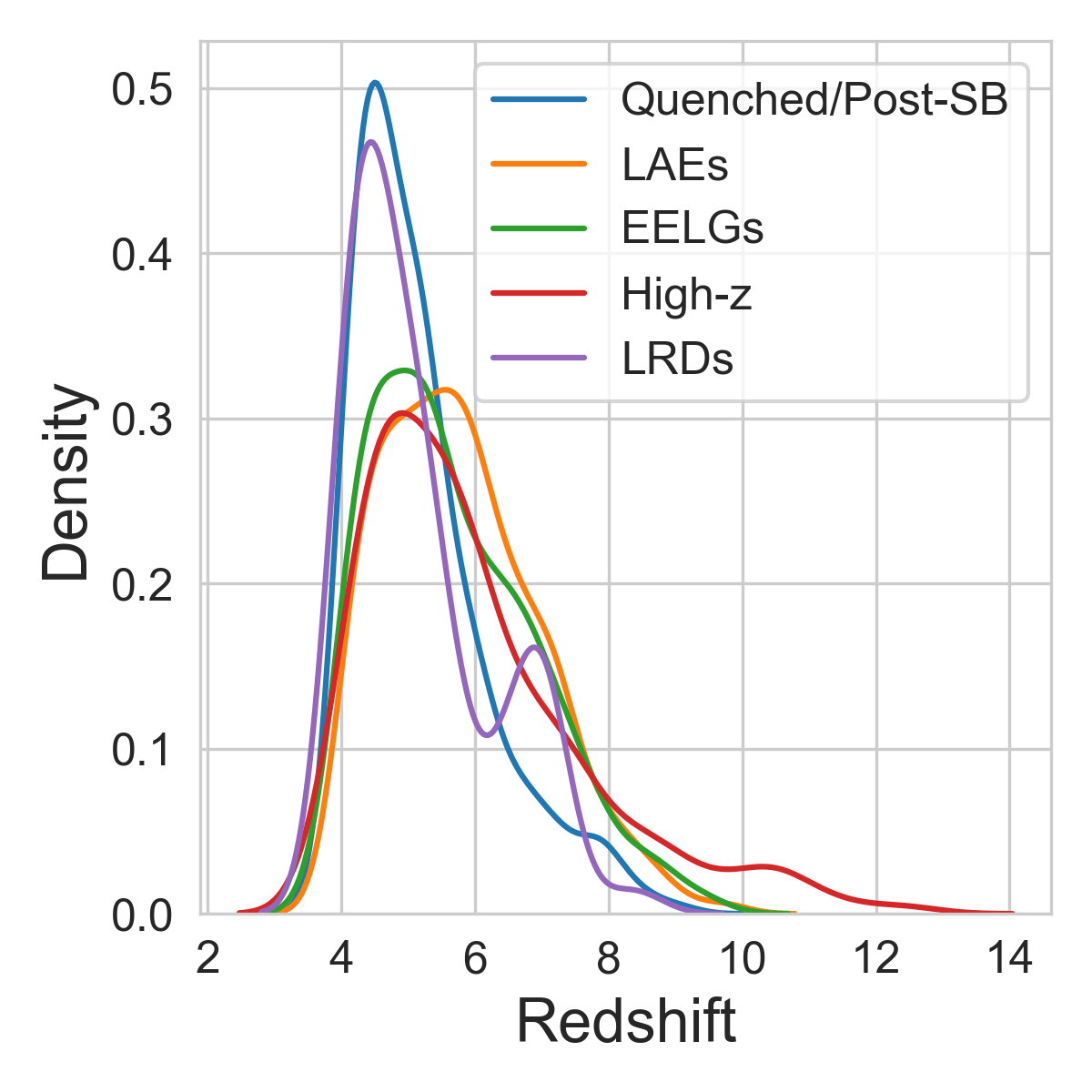}
    \vspace{-5mm}
    \caption{\textit{Left:} Observed, median-combined spectra of five exciting high redshift galaxy types identified using our VAE and clustering approach. The diversity of the galaxy spectra demonstrates the various physical processes that shape the continua and emission lines, enabling insights into galaxy evolution. \textit{Right:} The redshift distribution of the clusters indicating a redshift correlation between galaxy types identified naturally by the VAE.}
    \vspace{-5mm}
    \label{fig:spectra}
\end{figure*}
To explore which kinds of astrophysical sources have been clustered together, we create a median spectrum for each cluster, and then compare the resulting properties of the galaxies in each cluster with known galaxy types in the literature, tracing rare and unique phenomena in the early Universe. These labels are effectively assigned using prior knowledge of the expected spectroscopic properties of known distant galaxies, which up until now have largely been visually classified. In this work, we focus on five exciting classes of objects and briefly describe the importance of each galaxy type below: 

\textbf{Quenched/Post-starburst (SB) galaxies:} We identify 326 galaxies that can be classified as being in their (mini-) quenched or post-starburst phase \citep[e.g.][]{looser2025}, which is when a galaxy has recently undergone a burst of star-formation and is currently in its `lull' phase. Above a redshift of 4, our newly discovered sample nearly doubles the number of known such galaxies. 

\textbf{Lyman-$\alpha$ Emitters (LAEs):} Characterized by their strong Lyman-$\alpha$ emission at rest-frame wavelength of $1216$\,\AA, LAEs trace intense star-formation. At $z>6$, a strong Lyman-$\alpha$ line emerges from regions of the Universe that have been `reionized' due to UV photons from young stars, charting the phase transition of the intergalactic medium from a completely neutral to ionized state within a billion years after the Big Bang \citep[e.g.][]{saxena2024}. We identify 213 strong LAEs, doubling the number of LAEs currently known at $z>4$ \citep{tang2024, jones2025}.

\textbf{Extreme Emission Line Galaxies (EELGs):} These galaxies are characterized by extremely strong emission lines, tracing some of the highest star-formation rates in the Universe, driven by young, massive stars forming in short bursts. We identify 180 EELGs, more than doubling the number of such galaxies currently known at these redshifts \citep[e.g.][]{boyett2024}.

\textbf{High-redshift (High-z):} Galaxies at the highest redshifts trace galaxy formation immediately after the Big Bang. These galaxies typically lack heavier elements as evidenced by the weaker emission lines in their spectra \citep[e.g.][]{carniani2024, naidu2025}. With individual spectra lacking significant signal to enable a robust analysis of the underlying stars and gas, our identification of 320 sources that exhibit properties similar to some of the first galaxies, including some of the highest redshift galaxies in our parent sample, significantly expands the sample statistics for studying their properties in detail.

\textbf{Little Red Dots (LRDs):} An exciting discovery made using \textit{JWST} has been that of a handful of so-called Little Red Dots, which are extremely compact galaxies with a puzzling `V-shaped' continuum and strong emission lines, tracing both star-formation and supermassive black hole activity \citep{matthee2024}. Current models are unable to self-consistently explain the observed properties of LRDs without invoking exotic astrophysical phenomena \citep[e.g.][]{naidu2025b}. Our new sample of 142 LRDs will enable detailed spectroscopic analyses along with robust model comparison for these puzzling objects.

The median-combined original input spectra for these classes of objects are shown in the left-hand panel of Figure \ref{fig:spectra}, with the right-hand panel showing the redshift distribution of each class. Although the signal-to-noise ratio of individual galaxies that make up these combined spectra vary, it is clear from the unweighted median-combined spectra that individual galaxies within each cluster exhibit highly correlated spectral shapes and properties.

\section{Future Work}
\label{sec:future}
Since the original draft of this paper, the number of galaxy spectra available in public archives has grown substantially. The logical next step for the framework introduced here would be to retrain the model on these larger datasets and re-identify clusters of interesting galaxy types. Larger datasets also increase the probability of finding truly anomalous galaxy spectra, thereby expanding the discovery space. Additionally, `truth' labels assigned from smaller training samples can be leveraged to further automate the isolation of these interesting galaxy types from larger datasets.

Further improvements could be made to to the clustering methodology. At present, we identify clusters using GMMs in the collapsed 2D UMAP representation of the 16D latent space. Experiments could be performed on the performance of clustering directly in the latent space, as well as by implementing other clustering algorithms (such as DBSCAN, OPTICS, or hierarchical clustering) on the UMAP representation. Additionally, given the nature of the input data, there are likely to be degeneracies between the latent space parameters as the same galaxy spectrum could theoretically belong to multiple classes of known objects. Exploring these degeneracies further could help make the identification of clusters more robust.

Inclusion of multi-modal data, such as 2D imaging (or photometry) from \textit{JWST}, or accompanying spectra with higher spectral resolution (albeit with over limited wavelength ranges) capable of resolving finer spectroscopic features could add significant value to the clustering power by adding focus on additional important galaxy features. This would require changes to the VAE architecture to account for the increased complexity of the input data.

\section{Conclusions}
\label{sec:conclusions}
In this work, we have leveraged a Variational Autoencoder (VAE) architecture combined with clustering algorithms deployed on 2D representations of the learned latent space to identify unique and exciting classes of distant galaxies from publicly available \textit{JWST} spectroscopic data. Our approach has yielded significantly increased the number of objects belonging to known classes of interesting galaxies tracing unique physical phenomena in the early Universe in a highly automatic fashion. Increased samples of distant galaxies with interesting physical properties are desperately needed to test and refine theories of star and black hole formation in some of the first galaxies that formed after the Big Bang. 

With publicly available \textit{JWST} spectroscopic datasets steadily growing, our model architecture and implementation enables training and deployment at scale, providing a vital tool for astronomers to automate the identification of known galaxy types from large datasets in addition to discovering unknown, anomalous spectra that may trace new astrophysical phenomena. Our model can potentially be integrated into existing \textit{JWST} spectroscopic data pipelines and repositories to rapidly speed up automatic classification of interesting and/or anomalous galaxy spectra. Our input (processed) dataset and code is publicly available\footnote{\url{https://github.com/aayush3009/learnspec}}

\bibliography{speclearn}

@ARTICLE{kingma2013,
       author = {{Kingma}, Diederik P and {Welling}, Max},
        title = "{Auto-Encoding Variational Bayes}",
      journal = {arXiv e-prints},
         year = 2013,
        month = dec,
          eid = {arXiv:1312.6114},
        pages = {arXiv:1312.6114},
          doi = {10.48550/arXiv.1312.6114},
archivePrefix = {arXiv},
       eprint = {1312.6114},
 primaryClass = {stat.ML},
       adsurl = {https://ui.adsabs.harvard.edu/abs/2013arXiv1312.6114K}
}

@ARTICLE{mcinnes2019,
       author = {{McInnes}, Leland and {Healy}, John and {Melville}, James},
        title = "{UMAP: Uniform Manifold Approximation and Projection for Dimension Reduction}",
      journal = {arXiv e-prints},
         year = 2018,
        month = feb,
          eid = {arXiv:1802.03426},
        pages = {arXiv:1802.03426},
          doi = {10.48550/arXiv.1802.03426},
archivePrefix = {arXiv},
       eprint = {1802.03426},
 primaryClass = {stat.ML},
       adsurl = {https://ui.adsabs.harvard.edu/abs/2018arXiv180203426M}
}

@ARTICLE{rousseeuw1987,
        author = {Rousseeuw, Peter},
        year = {1987},
        month = {11},
        pages = {53-65},
        title = {Rousseeuw, P.J.: Silhouettes: A Graphical Aid to the Interpretation and Validation of Cluster Analysis. Comput. Appl. Math. 20, 53-65},
        volume = {20},
        journal = {Journal of Computational and Applied Mathematics},
        doi = {10.1016/0377-0427(87)90125-7}
}

@ARTICLE{looser2025,
       author = {{Looser}, Tobias J. and {D'Eugenio}, Francesco and {Maiolino}, Roberto and {Tacchella}, Sandro and {Curti}, Mirko and {Arribas}, Santiago and {Baker}, William M. and {Baum}, Stefi and {Bonaventura}, Nina and {Boyett}, Kristan and {Bunker}, Andrew J. and {Carniani}, Stefano and {Charlot}, Stephane and {Chevallard}, Jacopo and {Curtis-Lake}, Emma and {Lola Danhaive}, A. and {Eisenstein}, Daniel J. and {de Graaff}, Anna and {Hainline}, Kevin and {Ji}, Zhiyuan and {Johnson}, Benjamin D. and {Kumari}, Nimisha and {Nelson}, Erica and {Parlanti}, Eleonora and {Rix}, Hans-Walter and {Robertson}, Brant and {Del Pino}, Bruno Rodr{\'\i}guez and {Sandles}, Lester and {Scholtz}, Jan and {Smit}, Renske and {Stark}, Daniel P. and {{\"U}bler}, Hannah and {Williams}, Christina C. and {Willott}, Chris and {Witstok}, Joris},
        title = "{JADES: Differing assembly histories of galaxies: Observational evidence for bursty star formation histories and (mini-)quenching in the first billion years of the Universe}",
      journal = {Astronomy \& Astrophysics},
         year = 2025,
        month = may,
       volume = {697},
          eid = {A88},
        pages = {A88},
          doi = {10.1051/0004-6361/202347102},
archivePrefix = {arXiv},
       eprint = {2306.02470},
 primaryClass = {astro-ph.GA},
       adsurl = {https://ui.adsabs.harvard.edu/abs/2025A&A...697A..88L}
}

@ARTICLE{saxena2024,
       author = {{Saxena}, Aayush and {Bunker}, Andrew J. and {Jones}, Gareth C. and {Stark}, Daniel P. and {Cameron}, Alex J. and {Witstok}, Joris and {Arribas}, Santiago and {Baker}, William M. and {Baum}, Stefi and {Bhatawdekar}, Rachana and {Bowler}, Rebecca and {Boyett}, Kristan and {Carniani}, Stefano and {Charlot}, Stephane and {Chevallard}, Jacopo and {Curti}, Mirko and {Curtis-Lake}, Emma and {Eisenstein}, Daniel J. and {Endsley}, Ryan and {Hainline}, Kevin and {Helton}, Jakob M. and {Johnson}, Benjamin D. and {Kumari}, Nimisha and {Looser}, Tobias J. and {Maiolino}, Roberto and {Rieke}, Marcia and {Rix}, Hans-Walter and {Robertson}, Brant E. and {Sandles}, Lester and {Simmonds}, Charlotte and {Smit}, Renske and {Tacchella}, Sandro and {Williams}, Christina C. and {Willmer}, Christopher N.~A. and {Willott}, Chris},
        title = "{JADES: The production and escape of ionizing photons from faint Lyman-alpha emitters in the epoch of reionization}",
      journal = {Astronomy \& Astrophysics},
         year = 2024,
        month = apr,
       volume = {684},
          eid = {A84},
        pages = {A84},
          doi = {10.1051/0004-6361/202347132},
archivePrefix = {arXiv},
       eprint = {2306.04536},
 primaryClass = {astro-ph.GA},
       adsurl = {https://ui.adsabs.harvard.edu/abs/2024A&A...684A..84S}
}

@ARTICLE{tang2024,
       author = {{Tang}, Mengtao and {Stark}, Daniel P. and {Topping}, Michael W. and {Mason}, Charlotte and {Ellis}, Richard S.},
        title = "{JWST/NIRSpec Observations of Lyman {\ensuremath{\alpha}} Emission in Star-forming Galaxies at 6.5 {\ensuremath{\lesssim}} z {\ensuremath{\lesssim}} 13}",
      journal = {The Astrophysical Journal},
         year = 2024,
        month = nov,
       volume = {975},
       number = {2},
          eid = {208},
        pages = {208},
          doi = {10.3847/1538-4357/ad7eb7},
archivePrefix = {arXiv},
       eprint = {2408.01507},
 primaryClass = {astro-ph.GA},
       adsurl = {https://ui.adsabs.harvard.edu/abs/2024ApJ...975..208T}
}

@ARTICLE{jones2025,
       author = {{Jones}, Gareth C. and {Bunker}, Andrew J. and {Saxena}, Aayush and {Arribas}, Santiago and {Bhatawdekar}, Rachana and {Boyett}, Kristan and {Cameron}, Alex J. and {Carniani}, Stefano and {Charlot}, Stephane and {Curtis-Lake}, Emma and {Hainline}, Kevin and {Johnson}, Benjamin D. and {Kumari}, Nimisha and {Maseda}, Michael V. and {Rix}, Hans-Walter and {Robertson}, Brant E. and {Tacchella}, Sandro and {{\"U}bler}, Hannah and {Williams}, Christina C. and {Willott}, Chris and {Witstok}, Joris and {Zhu}, Yongda},
        title = "{JADES: measuring reionization properties using Lyman-alpha emission}",
      journal = {Monthly Notices of the Royal Astronomical Society},
         year = 2025,
        month = jan,
       volume = {536},
       number = {3},
        pages = {2355-2380},
          doi = {10.1093/mnras/stae2670},
archivePrefix = {arXiv},
       eprint = {2409.06405},
 primaryClass = {astro-ph.GA},
       adsurl = {https://ui.adsabs.harvard.edu/abs/2025MNRAS.536.2355J}
}

@ARTICLE{boyett2024,
       author = {{Boyett}, Kit and {Bunker}, Andrew J. and {Curtis-Lake}, Emma and {Chevallard}, Jacopo and {Cameron}, Alex J. and {Jones}, Gareth C. and {Saxena}, Aayush and {Charlot}, St{\'e}phane and {Curti}, Mirko and {Wallace}, Imaan E.~B. and {Arribas}, Santiago and {Carniani}, Stefano and {Willott}, Chris and {Alberts}, Stacey and {Eisenstein}, Daniel J. and {Hainline}, Kevin and {Hausen}, Ryan and {Johnson}, Benjamin D. and {Rieke}, Marcia and {Robertson}, Brant and {Stark}, Daniel P. and {Tacchella}, Sandro and {Williams}, Christina C. and {Chen}, Zuyi and {Egami}, Eiichi and {Endsley}, Ryan and {Kumari}, Nimisha and {Laseter}, Isaac and {Looser}, Tobias J. and {Maseda}, Michael V. and {Scholtz}, Jan and {Shivaei}, Irene and {Simmonds}, Charlotte and {Smit}, Renske and {{\"U}bler}, Hannah and {Witstok}, Joris},
        title = "{Extreme emission line galaxies detected in JADES JWST/NIRSpec - I. Inferred galaxy properties}",
      journal = {Monthly Notices of the Royal Astronomical Society},
         year = 2024,
        month = dec,
       volume = {535},
       number = {2},
        pages = {1796-1828},
          doi = {10.1093/mnras/stae2430},
archivePrefix = {arXiv},
       eprint = {2401.16934},
 primaryClass = {astro-ph.GA},
       adsurl = {https://ui.adsabs.harvard.edu/abs/2024MNRAS.535.1796B}
}

@ARTICLE{carniani2024,
       author = {{Carniani}, Stefano and {Hainline}, Kevin and {D'Eugenio}, Francesco and {Eisenstein}, Daniel J. and {Jakobsen}, Peter and {Witstok}, Joris and {Johnson}, Benjamin D. and {Chevallard}, Jacopo and {Maiolino}, Roberto and {Helton}, Jakob M. and {Willott}, Chris and {Robertson}, Brant and {Alberts}, Stacey and {Arribas}, Santiago and {Baker}, William M. and {Bhatawdekar}, Rachana and {Boyett}, Kristan and {Bunker}, Andrew J. and {Cameron}, Alex J. and {Cargile}, Phillip A. and {Charlot}, St{\'e}phane and {Curti}, Mirko and {Curtis-Lake}, Emma and {Egami}, Eiichi and {Giardino}, Giovanna and {Isaak}, Kate and {Ji}, Zhiyuan and {Jones}, Gareth C. and {Kumari}, Nimisha and {Maseda}, Michael V. and {Parlanti}, Eleonora and {P{\'e}rez-Gonz{\'a}lez}, Pablo G. and {Rawle}, Tim and {Rieke}, George and {Rieke}, Marcia and {Del Pino}, Bruno Rodr{\'\i}guez and {Saxena}, Aayush and {Scholtz}, Jan and {Smit}, Renske and {Sun}, Fengwu and {Tacchella}, Sandro and {{\"U}bler}, Hannah and {Venturi}, Giacomo and {Williams}, Christina C. and {Willmer}, Christopher N.~A.},
        title = "{Spectroscopic confirmation of two luminous galaxies at a redshift of 14}",
      journal = {Nature},
         year = 2024,
        month = sep,
       volume = {633},
       number = {8029},
        pages = {318-322},
          doi = {10.1038/s41586-024-07860-9},
archivePrefix = {arXiv},
       eprint = {2405.18485},
 primaryClass = {astro-ph.GA},
       adsurl = {https://ui.adsabs.harvard.edu/abs/2024Natur.633..318C}
}

@ARTICLE{naidu2025,
       author = {{Naidu}, Rohan P. and {Oesch}, Pascal A. and {Brammer}, Gabriel and {Weibel}, Andrea and {Li}, Yijia and {Matthee}, Jorryt and {Chisholm}, John and {Pollock}, Clara L. and {Heintz}, Kasper E. and {Johnson}, Benjamin D. and {Shen}, Xuejian and {Hviding}, Raphael E. and {Leja}, Joel and {Tacchella}, Sandro and {Ganguly}, Arpita and {Witten}, Callum and {Atek}, Hakim and {Belli}, Sirio and {Bose}, Sownak and {Bouwens}, Rychard and {Dayal}, Pratika and {Decarli}, Roberto and {de Graaff}, Anna and {Fudamoto}, Yoshinobu and {Giovinazzo}, Emma and {Greene}, Jenny E. and {Illingworth}, Garth and {Inoue}, Akio K. and {Kane}, Sarah G. and {Labbe}, Ivo and {Leonova}, Ecaterina and {Marques-Chaves}, Rui and {Meyer}, Romain A. and {Nelson}, Erica J. and {Roberts-Borsani}, Guido and {Schaerer}, Daniel and {Simcoe}, Robert A. and {Stefanon}, Mauro and {Sugahara}, Yuma and {Toft}, Sune and {van der Wel}, Arjen and {van Dokkum}, Pieter and {Walter}, Fabian and {Watson}, Darach and {Weaver}, John R. and {Whitaker}, Katherine E.},
        title = "{A Cosmic Miracle: A Remarkably Luminous Galaxy at $z_{\rm{spec}}=14.44$ Confirmed with JWST}",
      journal = {arXiv e-prints},
         year = 2025,
        month = may,
          eid = {arXiv:2505.11263},
        pages = {arXiv:2505.11263},
          doi = {10.48550/arXiv.2505.11263},
archivePrefix = {arXiv},
       eprint = {2505.11263},
 primaryClass = {astro-ph.GA},
       adsurl = {https://ui.adsabs.harvard.edu/abs/2025arXiv250511263N}
}

@ARTICLE{matthee2024,
       author = {{Matthee}, Jorryt and {Naidu}, Rohan P. and {Brammer}, Gabriel and {Chisholm}, John and {Eilers}, Anna-Christina and {Goulding}, Andy and {Greene}, Jenny and {Kashino}, Daichi and {Labbe}, Ivo and {Lilly}, Simon J. and {Mackenzie}, Ruari and {Oesch}, Pascal A. and {Weibel}, Andrea and {Wuyts}, Stijn and {Xiao}, Mengyuan and {Bordoloi}, Rongmon and {Bouwens}, Rychard and {van Dokkum}, Pieter and {Illingworth}, Garth and {Kramarenko}, Ivan and {Maseda}, Michael V. and {Mason}, Charlotte and {Meyer}, Romain A. and {Nelson}, Erica J. and {Reddy}, Naveen A. and {Shivaei}, Irene and {Simcoe}, Robert A. and {Yue}, Minghao},
        title = "{Little Red Dots: An Abundant Population of Faint Active Galactic Nuclei at z {\ensuremath{\sim}} 5 Revealed by the EIGER and FRESCO JWST Surveys}",
      journal = {The Astrophysical Journal},
         year = 2024,
        month = mar,
       volume = {963},
       number = {2},
          eid = {129},
        pages = {129},
          doi = {10.3847/1538-4357/ad2345},
archivePrefix = {arXiv},
       eprint = {2306.05448},
 primaryClass = {astro-ph.GA},
       adsurl = {https://ui.adsabs.harvard.edu/abs/2024ApJ...963..129M}
}

@ARTICLE{naidu2025b,
       author = {{Naidu}, Rohan P. and {Matthee}, Jorryt and {Katz}, Harley and {de Graaff}, Anna and {Oesch}, Pascal and {Smith}, Aaron and {Greene}, Jenny E. and {Brammer}, Gabriel and {Weibel}, Andrea and {Hviding}, Raphael and {Chisholm}, John and {Labb\textbackslash'e}, Ivo and {Simcoe}, Robert A. and {Witten}, Callum and {Atek}, Hakim and {Baggen}, Josephine F.~W. and {Belli}, Sirio and {Bezanson}, Rachel and {Boogaard}, Leindert A. and {Bose}, Sownak and {Covelo-Paz}, Alba and {Dayal}, Pratika and {Fudamoto}, Yoshinobu and {Furtak}, Lukas J. and {Giovinazzo}, Emma and {Goulding}, Andy and {Gronke}, Max and {Heintz}, Kasper E. and {Hirschmann}, Michaela and {Illingworth}, Garth and {Inoue}, Akio K. and {Johnson}, Benjamin D. and {Leja}, Joel and {Leonova}, Ecaterina and {McConachie}, Ian and {Maseda}, Michael V. and {Natarajan}, Priyamvada and {Nelson}, Erica and {Setton}, David J. and {Shivaei}, Irene and {Sobral}, David and {Stefanon}, Mauro and {Tacchella}, Sandro and {Toft}, Sune and {Torralba}, Alberto and {van Dokkum}, Pieter and {van der Wel}, Arjen and {Volonteri}, Marta and {Walter}, Fabian and {Wang}, Bingjie and {Watson}, Darach},
        title = "{A ``Black Hole Star'' Reveals the Remarkable Gas-Enshrouded Hearts of the Little Red Dots}",
      journal = {arXiv e-prints},
         year = 2025,
        month = mar,
          eid = {arXiv:2503.16596},
        pages = {arXiv:2503.16596},
          doi = {10.48550/arXiv.2503.16596},
archivePrefix = {arXiv},
       eprint = {2503.16596},
 primaryClass = {astro-ph.GA},
       adsurl = {https://ui.adsabs.harvard.edu/abs/2025arXiv250316596N}
}

@ARTICLE{nicolau2025,
       author = {{Nicolaou}, C. and {Nathan}, R.~P. and {Lahav}, O. and {Palmese}, A. and {Saintonge}, A. and {Aguilar}, J. and {Ahlen}, S. and {Allende Prieto}, C. and {Bailey}, S. and {BenZvi}, S. and {Bianchi}, D. and {Brodzeller}, A. and {Brooks}, D. and {Claybaugh}, T. and {de la Macorra}, A. and {Della Costa}, J. and {Dey}, Arjun and {Doel}, P. and {Forero-Romero}, J.~E. and {Gazta{\~n}aga}, E. and {Gontcho}, S. Gontcho A and {Gutierrez}, G. and {Honscheid}, K. and {Howlett}, C. and {Ishak}, M. and {Kehoe}, R. and {Kirkby}, D. and {Kisner}, T. and {Kremin}, A. and {Lambert}, A. and {Landriau}, M. and {Le Guillou}, L. and {Meisner}, A. and {Miquel}, R. and {Moustakas}, J. and {Nadathur}, S. and {Prada}, F. and {P{\'e}rez-R{\`a}fols}, I. and {Rossi}, G. and {Sanchez}, E. and {Schubnell}, M. and {Siudek}, M. and {Sprayberry}, D. and {Tarl{\'e}}, G. and {Weaver}, B.~A. and {Zou}, H.},
        title = "{Identifying Anomalous DESI Galaxy Spectra with a Variational Autoencoder}",
      journal = {arXiv e-prints},
         year = 2025,
        month = jun,
          eid = {arXiv:2506.17376},
        pages = {arXiv:2506.17376},
          doi = {10.48550/arXiv.2506.17376},
archivePrefix = {arXiv},
       eprint = {2506.17376},
 primaryClass = {astro-ph.IM},
       adsurl = {https://ui.adsabs.harvard.edu/abs/2025arXiv250617376N}
}

@ARTICLE{portillo2020,
       author = {{Portillo}, Stephen K.~N. and {Parejko}, John K. and {Vergara}, Jorge R. and {Connolly}, Andrew J.},
        title = "{Dimensionality Reduction of SDSS Spectra with Variational Autoencoders}",
      journal = {The Astronomical Journal},
         year = 2020,
        month = jul,
       volume = {160},
       number = {1},
          eid = {45},
        pages = {45},
          doi = {10.3847/1538-3881/ab9644},
archivePrefix = {arXiv},
       eprint = {2002.10464},
 primaryClass = {astro-ph.IM},
       adsurl = {https://ui.adsabs.harvard.edu/abs/2020AJ....160...45P}
}

@ARTICLE{bohm2023,
       author = {{B{\"o}hm}, Vanessa and {Kim}, Alex G. and {Juneau}, St{\'e}phanie},
        title = "{Fast and efficient identification of anomalous galaxy spectra with neural density estimation}",
      journal = {Monthly Notices of the Royal Astronomical Society},
         year = 2023,
        month = dec,
       volume = {526},
       number = {2},
        pages = {3072-3087},
          doi = {10.1093/mnras/stad2773},
archivePrefix = {arXiv},
       eprint = {2308.00752},
 primaryClass = {astro-ph.IM},
       adsurl = {https://ui.adsabs.harvard.edu/abs/2023MNRAS.526.3072B}
}

@ARTICLE{scourfield2023,
       author = {{Scourfield}, M. and {Saintonge}, A. and {de Mijolla}, D. and {Viti}, S.},
        title = "{De-noising of galaxy optical spectra with autoencoders}",
      journal = {Monthly Notices of the Royal Astronomical Society},
         year = 2023,
        month = dec,
       volume = {526},
       number = {2},
        pages = {3037-3050},
          doi = {10.1093/mnras/stad2709},
archivePrefix = {arXiv},
       eprint = {2309.02315},
 primaryClass = {astro-ph.GA},
       adsurl = {https://ui.adsabs.harvard.edu/abs/2023MNRAS.526.3037S}
}



\end{document}